\documentclass{emulateapj}
\usepackage{epsf}
\usepackage{apjfonts}
\usepackage{xspace}
\usepackage{amsmath}
\usepackage{framed}
\def\figdir{.}

\def\H2{{{\rm H}_2}}
\def\HI{{\rm H\,I}}

\def\Msun{\, {\rm M}_{\odot}}

\def\Sgas{\Sigma_{\rm H}}
\def\Smol{\Sigma_{\H2}}
\def\Shi{\Sigma_{\HI}}
\def\Ssfr{\Sigma_{\rm SFR}}

\def\dim#1{\mbox{\,#1}}
\def\figname#1#2{\figdir/#1}

\def\hide#1{}

\begin{document}

\title{On the Kennicutt-Schmidt relation of low-metallicity high-redshift galaxies}

\author{Nickolay Y.\
  Gnedin\altaffilmark{1,2,3} and Andrey
  V. Kravtsov\altaffilmark{2,3,4}}  
\altaffiltext{1}{Particle Astrophysics Center, 
Fermi National Accelerator Laboratory, Batavia, IL 60510, USA; gnedin@fnal.gov}
\altaffiltext{2}{Department of Astronomy \& Astrophysics, The
  University of Chicago, Chicago, IL 60637 USA} 
\altaffiltext{3}{Kavli Institute for Cosmological Physics and Enrico
  Fermi Institute, The University of Chicago, Chicago, IL 60637 USA;
  andrey@oddjob.uchicago.edu} 
\altaffiltext{4}{Enrico Fermi Institute, The University of Chicago,
Chicago, IL 60637}

\begin{abstract}
We present results of self-consistent, high-resolution cosmological
simulations of galaxy formation at $z\sim 3$. The simulations employ
recently developed recipe for star formation based on the local
abundance of molecular hydrogen, which is tracked self-consistently
during the course of simulation. The phenomenological $\H2$ formation
model accounts for the effects of dissociating UV radiation of stars
in each galaxy, as well as self-shielding and shielding of $\H2$ by
dust, and therefore allows us to explore effects of lower
metallicities and higher UV fluxes prevalent in high redshift galaxies
on their star formation. We compare stellar masses, metallicities, and
star formation rates of the simulated galaxies to available
observations of the Lyman Break Galaxies (LBGs) and find a reasonable
agreement.  We find that the Kennicutt-Schmidt (KS) relation exhibited
by our simulated galaxies at $z\approx 3$ is substantially steeper and
has a lower amplitude than the $z=0$ relation at $\Sgas\la 100\rm\
M_{\odot}\,pc^{-2}$. The predicted relation, however, is consistent
with existing observational constraints for the $z\approx 3$ Damped
Lyman $\alpha$ (DLA) and LBGs.  Our tests show that the main reason
for the difference from the local KS relation is lower metallicity of
the ISM in high redshift galaxies. We discuss several implications of
the metallicity-dependence of the KS relation for galaxy evolution and
interpretation of observations. In particular, we show that the observed size of high-redshift exponential disks
depends sensitively on their KS relation. Our results also suggest that
significantly reduced star formation efficiency at low gas surface
densities can lead to strong suppression of star formation in low-mass
high-redshift galaxies and long gas consumption time scales over most
of the disk in large galaxies.  The longer gas consumption time scales could make disks
more resilient to major and minor mergers and could help explain the
prevalence of the thin stellar disks in the local universe.
\end{abstract}

\keywords{cosmology: theory -- galaxies: evolution -- galaxies:
  formation -- stars:formation -- methods: numerical}

\section{Introduction}
\label{sec:intro}

The hierarchical Cold Dark Matter (CDM) structure formation paradigm
\citep{white_rees78,blumenthal_etal84} has proved to be remarkably
successful both in predicting and explaining a variety of
observational data from the detailed properties of the temperature
anisotropies of the Cosmic Microwave Background \citep{dunkley_etal09}
to the linear \citep[e.g.,][]{tegmark_etal06} and nonlinear
\citep[e.g.,][]{cosmo:sfw06,conroy_etal06,kravtsov06,wetzel_white09}
clustering of galaxies.  

Galaxy formation modeling within the CDM framework has produced a
general picture of how galaxy-sized objects collapse out of primordial
density fluctuations
\citep[e.g.,][]{steinmetz_navarro02,misc:sh03,mayer_etal08}. Nevertheless, some
key aspects of the picture are still being developed
\citep{birnboim_dekel03,keres_etal05} and accurate predictions of even
the basic properties of galaxies, such as luminosities, colors,
metallicities, and gas content remain a challenge
\citep[e.g.,][]{mayer_etal08}. These properties depend sensitively on
the details of star formation. The root of the
challenge is therefore our lack of full understanding of
how gas is converted into stars under different conditions during
different stages of galaxy evolution.

Empirical studies of the Kennicutt-Schmidt (KS) relation
\citep{schmidt59,sfr:k89,sfr:k98a} between the total surface density
of the neutral gas $\Sgas$, or its specific phase (atomic or
molecular), and surface density of star formation rate $\Ssfr$
established that the KS relation can be approximated by a power-law,
$\Ssfr\propto \Sgas^{n}$, with $n \approx 1.4$ (or, more generally,
between 1 and 2) at high surface densities, when $\Sgas$ is dominated
by the molecular gas, but steepens considerably at at $\Sgas \la
10\Msun/\dim{pc}^2$
\citep{sfr:k98a,sfr:mk01,misc:wb02,heyer_etal04,boissier_etal07,sfr:blwb08,roychowdhury_etal09,bolatto_etal09}. These
studies, however, have been limited to samples of nearby galaxies with
a relatively narrow range of metallicities ($\approx 0.5-1Z_{\odot}$)
and morphologies \citep[thin gaseous disks; see
however][]{roychowdhury_etal09}.

Observational picture remains sketchy at higher redshifts and lower
metallicities. On one hand, recent studies show that the power law
KS relation between star formation and {\it molecular\/} gas surface
density holds in actively star forming galaxies at higher redshifts
\citep[e.g.,][]{sfr:b07,baker_etal04a}. On the other hand,
recent searches for the UV emission associated with young stars in
damped Ly$\alpha$ systems \citep[DLA; ][]{sfr:wc06} and in Ca\,II
absorbers at low redshifts \citep{wild_etal07} indicate that $\Ssfr$
at the gas surface densities characteristic for these systems
($N_{\HI}\approx 10^{21}-10^{22}\dim{cm}^{-2}$ or $\Sgas \approx 10 -
100\Msun/\dim{pc}^2$) is an order of magnitude lower than implied by
the local KS relation. A similar result is indicated by recent studies
of the Lyman Break Galaxies \citep{hizgal:rwcc09,rafelski09}.

Given that the typical metallicity of the DLA systems is considerably
lower than solar\footnote{We refer to ``solar metallicity'' as a
typical gas metallicity in the solar neighborhood, $Z_{\odot}=0.019$
by mass, rather than the metallicity of the Sun, which has been
recently revised downward.}  \citep[][and referencse
therein]{igm:fplp09}, this may indicate that the $\Ssfr-\Sgas$
relation is significantly lower at these surface densities in systems
of low metallicity. The first evidence of this in the nearby galaxies
is that the KS relation in the SMC appears to be significantly steeper
than $n \approx 1.4$ at $\Sgas\la50\Msun/\dim{pc}^2$
\citep{bolatto_etal09}.

Overall, the trends described above can be expected.  Observations
show that star formation in galaxies correlates strongly with the {\it
molecular\/} gas \citep[e.g.,][]{wong_blitz02}. At the same time, SFR
has only weak or no correlation with the density of atomic gas
\citep{wong_blitz02,kennicutt_etal07,sfr:blwb08}.  We can thus expect
that $\Ssfr$ relation with $\Sgas=\Smol+\Shi$ will depend on the
molecular fraction of the gas $f_\H2 \equiv \Smol/\Sgas$. The
molecular fraction in observed galaxies is controlled by the gas
density, FUV flux, and pressure \citep[e.g.,][]{blitz_rosolowsky06}, consistent
with theoretical expectations
\citep{elmegreen93,elmegreen_parravano94,robertson_kravtsov08}, but is
also expected to be sensitive to the dust content (and thus
metallicity) of the gas because dust plays an important role both in
shielding molecular gas from UV radiation and in catalyzing production
of $\H2$ \citep[e.g.,][]{stahler_palla05}. Moreover, the efficiency
with which a galaxy is able to build high-density molecular regions
should depend on metallicity, because the compression of gas in the
radiative shocks, arising in the highly turbulent interstellar medium
(ISM) of gaseous disks, should depend on the metallicity-dependent
cooling rate.

These considerations suggest that in order to make progress in our
modeling of star formation in galaxies, we need simulations with both
the high spatial and mass resolution to resolve the density structure
and thermal state of different regions of the ISM in galaxies and a
model for formation of molecular hydrogen.  Several theoretical models
incorporating some of these elements and potentially capable of
predicting dependence of star formation relation on ISM properties
have been developed recently
\citep{sfr:km05,pelupessy_etal06,robertson_kravtsov08,tasker_bryan08,schaye_dallavecchia08,ng:gtk09,tasker_tan09,krumholz_etal09,pelupessy_popadopoulos09}.
These models have shown that thermodynamics of the ISM can control the
shape of the density PDF of the ISM gas
\citep{wada_norman01,wada_norman07,wada_etal02,kravtsov03,tasker_bryan08,robertson_kravtsov08}
which, in turn, can affect the KS relation. The general prediction of
the models that incorporate the physics of atomic-to-molecular
transition at the boundaries of molecular clouds is that the density
of this transition, controlled primarily by dust abundance (i.e.,
metallicity of the gas), is a key factor affecting the amplitude and
slope of the KS relation \citep{ng:gtk09,krumholz_etal09}.  The models
of \citet{robertson_kravtsov08} have also highlighted importance of
the interstellar radiation field for controlling the abundance of
low-density, diffuse molecular gas \citep{elmegreen93}, as well as the
slope and amplitude of the large-scale KS relation and metallicity of
the gas. So far, however, none of the models have considered ISM of
galaxies arising and evolving self-consistently in cosmological
context.

In this paper, we present results of self-consistent, high-resolution
cosmological simulations of galaxy formation, which employ the
metallicity-dependent model of $\H2$ presented in \citet*{ng:gtk09}
and \citet*{ng:gk10b}. In the latter paper we also study sensitivity
of global star formation to the full range of metallicities and UV
fluxes.  The focus of this paper is to explore the predictions of our
$\H2$-based star formation recipe for the KS relation in the
low-metallicity, high UV-flux environments of high-redshift galaxies
and compare these predictions with the existing observational
constraints from DLA and Lyman Break Galaxies (LBGs).

\section{Simulations and star formation model}
\label{sec:sims}

The physical ingredients and computational setup for the simulation we
use in this paper are described in \citet{ng:gtk09}. As a brief
summary, the Adaptive Refinement Tree (ART) code \citep{kravtsov99,kravtsov_etal02,rudd_etal08}
uses adaptive mesh refinement in both the gas dynamics and gravity
calculations to achieve high dynamic range in spatial scale. The
high-mass resolution is achieved by using the standard ``zoom-in''
initial conditions with particles of smaller mass resolving Lagrangian
region of a system of interest.

The specific initial conditions setup we use in this paper is a
Lagrangian region corresponding to five virial radii of a galactic
system, which evolves to approximately Milky Way mass ($M\approx
10^{12}\Msun$) at $z=0$, with the mass resolution of
$1.3\times10^6\Msun$ in dark matter, $2.2\times 10^5\Msun$ in baryons,
and with the spatial resolution of
$65\dim{pc}\times[4/(1+z)]$ (in physical units) within the
region. This Lagrangian region is embedded into a cubic volume of
$6h^{-1}$ comoving Mpc on a side, which is followed with a coarse
$64^3$ grid outside the Lagrangian region with periodic boundary
conditions.

Our fiducial simulation includes star formation and supernova enrichment and
thermal energy feedback, and follows self-consistently the 3D
radiative transfer of UV radiation from individual stellar particles
using the OTVET approximation \citep{ng:ga01}. The simulation
incorporates non-equilibrium chemical network of hydrogen and helium
and non-equilibrium, metallicity-dependent cooling and heating rates, which make use of the
local abundance of atomic, molecular, and ionic species and UV
intensity, followed self-consistently during the course of the
simulation. 

A phenomenological model of molecular hydrogen formation is used to
identify the locations of molecular clouds in the simulated galaxies,
as described in \citet{ng:gtk09} and, in more detail in
\citet{ng:gk10b}. The model incorporates both self-shielding of
molecular hydrogen from the UV radiation and the shielding by dust and
takes into account catalyzing effect of dust on $\H2$ formation. The
local abundance of dust in our model is assumed to linearly scale with
gas metallicity, which is consistent with observations of the MW, LMC,
and SMC \citep{ng:gtk09,ng:gk10b}. Gas metallicity in the simulations
is modeled by including metal enrichment by supernovae (both type II
and Ia) assuming standard yields, and by stars via wind mass loss,
with the total amount of heavy elements released into the ISM
determined by the assumed Miller-Scalo IMF of the stars \cite[see][for
details]{kravtsov03}. Once released into the gas the heavy elements
are advected by the code in the same way as the gas.

We calibrate our $\H2$ model to fit the observed atomic and molecular
gas fraction as a function of column density in the different
metallicity environments of the Milky Way, LMC, and SMC and other
nearby galaxies \citep{ng:gtk09,ng:gk10b}. The model reproduces the
metallicity dependence of the column density at which a sharp
transition from the atomic to fully molecular gas occurs. 

The star formation recipe adopted in this work is similar to the
recipe 2 of \citet{ng:gtk09}, albeit with somewhat different parameters. Specifically, stellar particles are formed in
cells that have mass fraction of molecular hydrogen higher than $f_{\rm H_2}\geq 0.1$ according to the rate:
\begin{equation}
 \dot{\rho}_\star =
\frac{\varepsilon_{\rm ff}}{\tau_{\rm sf}} \rho_{\H2}, 
\label{eq:sfr}
\end{equation}
where $\rho_{\rm H_2}$ is the density of molecular hydrogen,
$\tau_{\rm sf} = \min(\tau_{\rm max},\tau_{\rm ff})$, $\tau_{\rm ff} =
\sqrt{3\pi/32G\rho}$ is the free-fall time for a uniform sphere,
$\tau_{\rm max}=6.65\times 10^6$~years is the free-fall time in the gas with the total
hydrogen number density of $50\dim{cm}^{-3}$, and $\varepsilon_{\rm
ff}$ is star formation efficiency per free fall time
\citep[e.g.,][]{sfr:km05,sfr:mo07}. We adopt the efficiency of
$\varepsilon_{\rm ff}=0.005$ consistent with observational constraints
for the average efficiency of star formation in the molecular clouds
of the Milky Way and other nearby galaxies \citep{sfr:kt07,sfr:mo07}.
The $\tau_{\rm sf}$ we adopt assumes that in low density cells, in
which molecular fraction $f_{\H2}$ is below unity, star formation proceeds
mainly in unresolved molecular clouds on subgrid scales. This
assumption then also motivates setting the maximum free fall time to
$\tau_{\rm max}$ corresponding to the number density of $50\,\,\rm
cm^{-3}$ typical average density of molecular clouds.  The
$f_{\H2}<1$ in these cells then can be viewed as reflecting the
fraction of the total gas in such star forming molecular clouds, which
themselves have $f_{\H2}=1$, rather than incomplete conversion of the
atomic gas into the molecular form inside the clouds.

Note that the empirical calibration of both the $\H2$ formation model
calibration and the efficiency of star formation is done on small
scales of individual molecular clouds.  The results presented in this
paper, however, are on global kiloparsec scales and the model was not
adjusted in any way to produce these results.

\section{Star formation in high redshift galaxies}
\label{sec:highz}

\subsection{Global properties of simulated galaxies}

\begin{figure}[t]
\epsscale{1.15}
\plotone{\figname{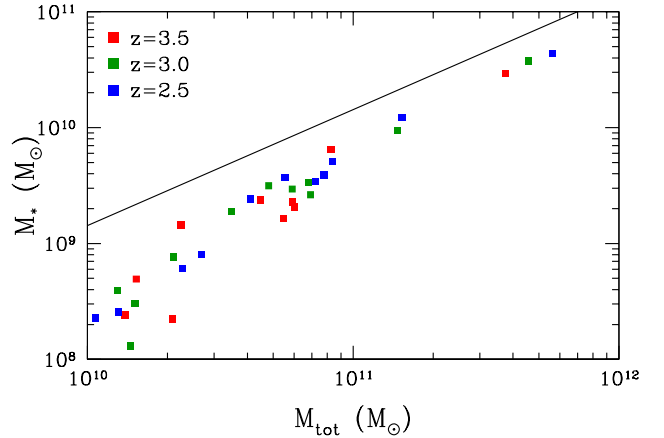}{f1.eps}}
\caption{The stellar mass of fully resolved model galaxies versus the total 
  mass of their halos.  The solid
  black line is the limit of the complete conversion of baryons into
  stars, $M_* = (\Omega_B/\Omega_M) M_{\rm tot}$. The sets of colored points
  correspond to the three redshifts in the simulation that cover the
  range of redshifts probed by observed systems of \citet{sfr:wc06}.}
\label{fig:masses}
\end{figure}

\begin{figure}[t]
\epsscale{1.15}
\plotone{\figname{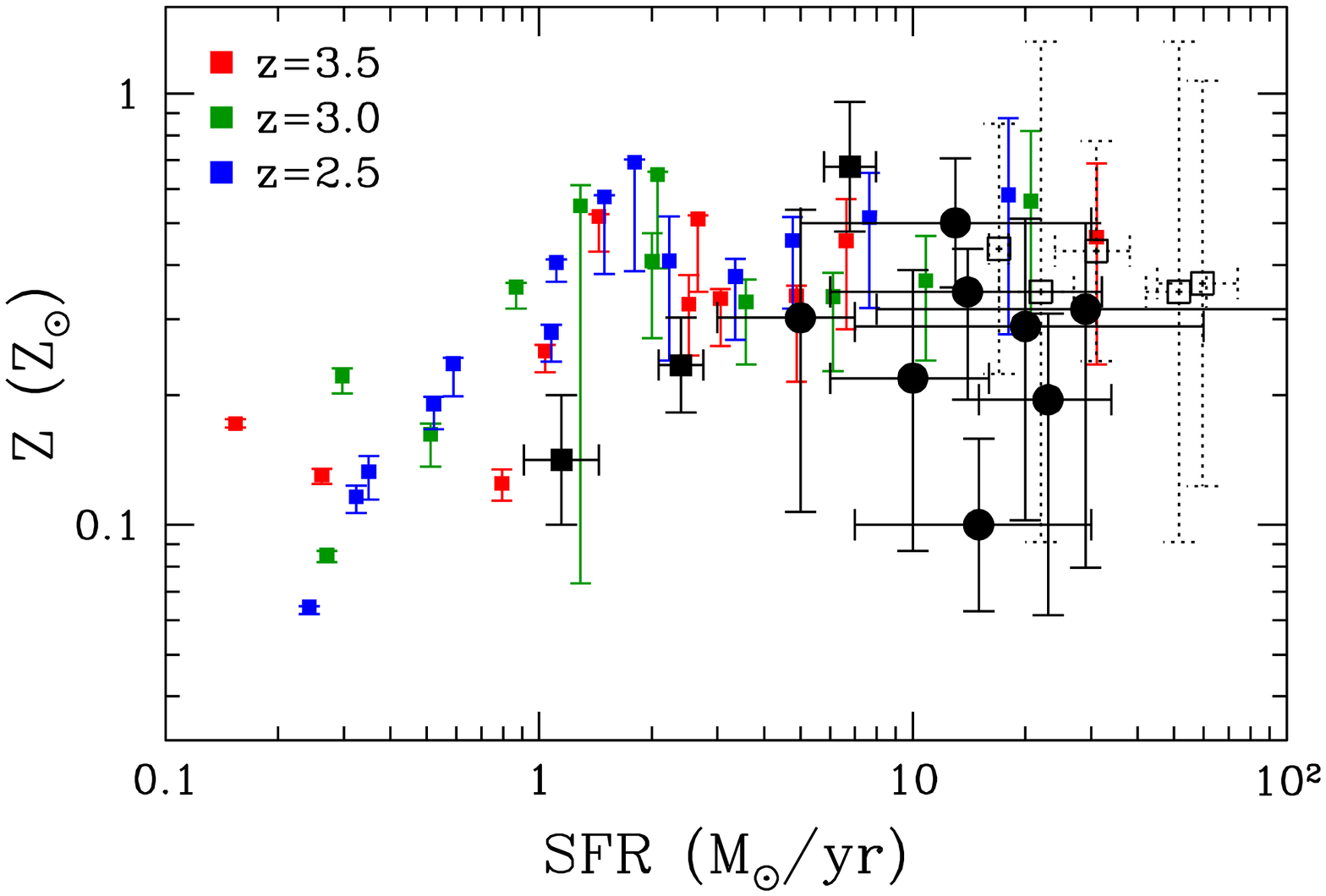}{f2a.eps}}
\plotone{\figname{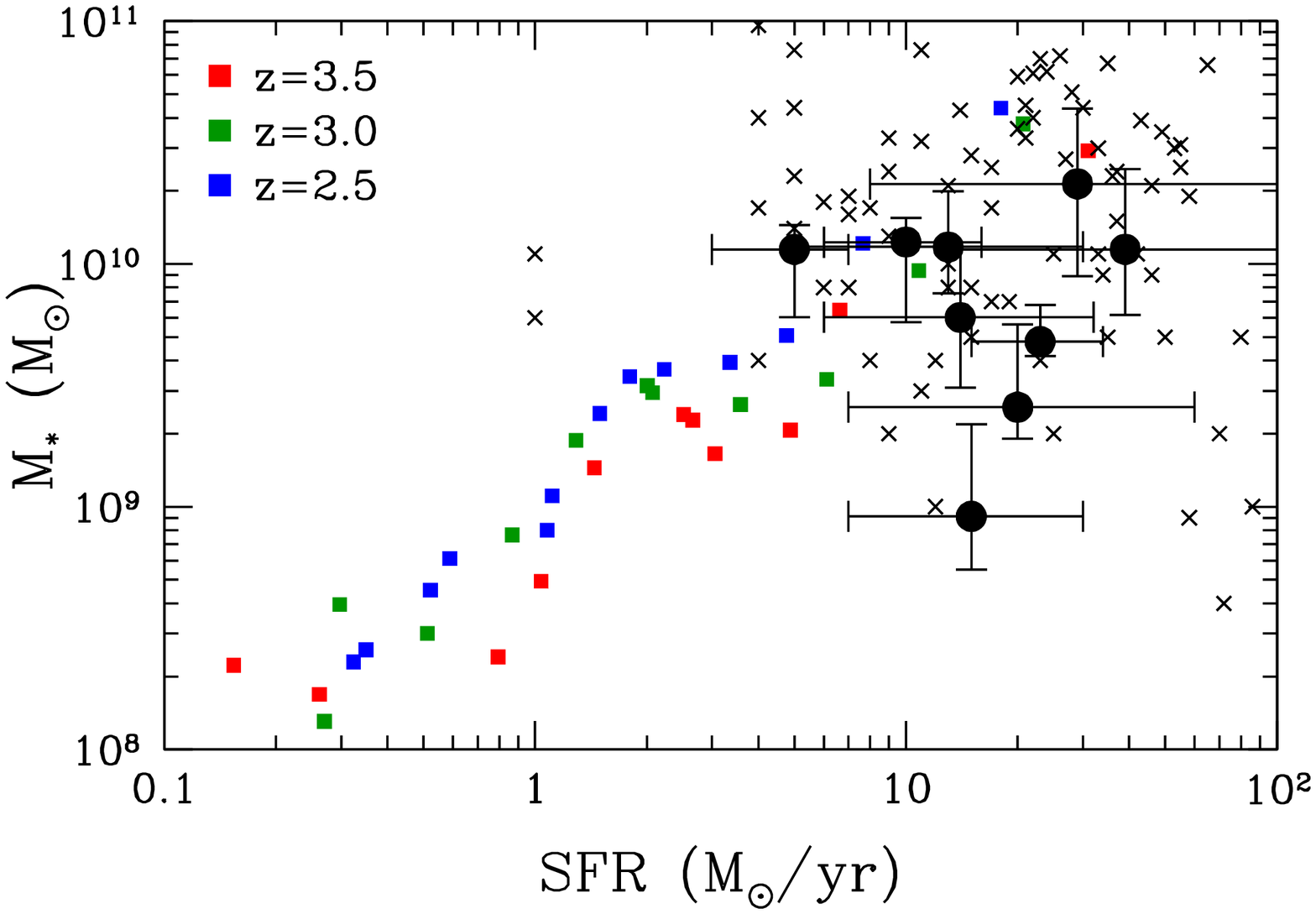}{f2b.eps}}
\plotone{\figname{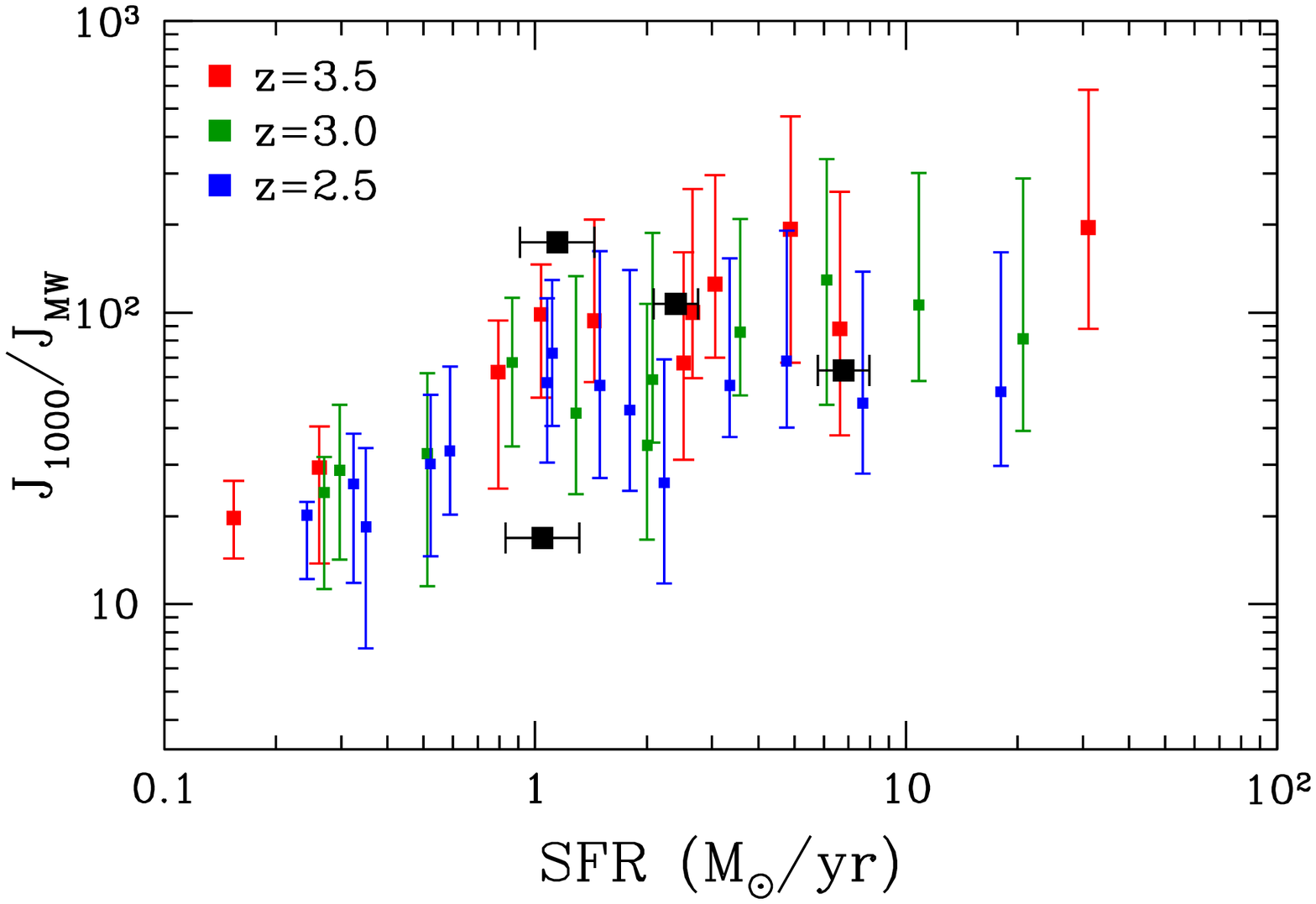}{f2c.eps}}
\caption{The gas metallicity (top), the stellar mass (middle), and the
  radiation field (bottom) of fully resolved model galaxies versus their star
  formation rate. The error bars on the metallicity and radiation
  field points show the 10\% - 90\% range of these properties measured
  in the individual 
  galaxies. The sets of colored points correspond to the three
  redshifts in the simulation that cover the range of redshifts probed
  by observed systems of \citet{sfr:wc06}. Black filled circles,
  crosses, and open squares are the observational measurements for LBG
  galaxies from \citet{hizgal:mcmm09}, \citet{hizgal:essp06}, and 
  \citet{hizgal:pssc01}, respectively. Black squares are measurements
  for GRB host galaxies from \citet{hizgal:cppp09}.}  
\label{fig:gals}
\end{figure}

Before we address the star formation relations, we need to consider
the bulk properties of simulated galaxies to make sure they can be
plausible counterparts of observed LBGs and DLAs.  In addition to the
main Milky Way progenitor, the fully resolved Lagrangian region of the
simulation volume contains a number of smaller mass
galaxies. Figure~\ref{fig:masses} shows the stellar masses of the
simulated galaxies versus the total mass of their parent halos,
defined as a mass within a spherical radius enclosing overdensity of
180 with respect to the mean density, at redshifts $2.5<z<3.5$.\footnote{We
only include in Fig.~\ref{fig:masses} and the following figures
galaxies whose centers are fully resolved to the highest resolution of
our simulation ($65\dim{pc}$ at $z=3$, nine levels of refinement),
with the sole exception of Fig.~\ref{fig:msmt}.} The figure shows that
simulated galaxies span almost two orders of magnitude of total mass
($M_{\rm tot}\sim 10^{10}-6\times 10^{11}\,\rm M_{\odot}$) and more than two orders of
magnitude of stellar mass ($M_{\ast}\sim 10^8-5\times 10^{10}\ \rm
M_{\odot}$). Interestingly, the $M_{\ast}-M_{\rm tot}$ correlation has
the shape qualitatively similar to that required to explain the
difference between the predicted halo mass function and observed
stellar mass function at lower redshifts
\citep[e.g.,][]{conroy_wechsler09,moster_etal09a}. This indicates that processes
included in our simulations could, at least partly, be responsible for
the observed shape of stellar mass function. We note, however, that
normalization of this relation in our simulations is considerably
higher than that indicated by observational analyzes \citep[e.g., see
Fig. 2 in][]{conroy_wechsler09}, especially for lower mass
systems. This indicates that star formation in small-mass halos is
still considerably more efficient in our simulations than in real
galaxies.

The star formation rates, stellar masses, metallicities, and radiation
fields of these galaxies are shown in Figures \ref{fig:gals} at three
redshifts in the range $2.5<z<3.5$.  For 
comparison, we also show for the observational estimates of the same
properties for the Lyman Break Galaxies from
\citet{hizgal:mcmm09},  \citet{hizgal:essp06}, and \citet{hizgal:pssc01}. Since the volume of
our simulation is relatively small, the overlap with the observed
galaxies in stellar mass is limited. In the region of overlap,
however, our model galaxies have star formation rates and stellar
masses consistent with those of observed LBGs.  The gas metallicities
of our model galaxies are consistent with observations within the
errorbars\footnote{We note, however, that metallicities in the
simulations carry all the systematic uncertainties of the adopted
nucleosynthetic yields and of the IMF.}. 

The average metallicities of the simulated galaxies are higher than
the typical metallicity of DLA systems at $z\sim 3$
\citep{prochaska_etal03,fynbo_etal08}. This may be because the masses
of most DLA systems are lower than the range of halo masses resolved
in our simulations or because DLA systems probe the outer regions of
the disks where metallicities are systematically lower due to
metallicity gradients. It is also possible that we overestimate star
formation efficiency and hence metal enrichment in objects of a given
mass. The trends of star formation relations with metallicity
discussed below can thus be expected to be even stronger for the real
DLA systems compared to the higher-metallicity galaxies in our
simulations.

The radiation field in a galaxy may vary widely between different
locations and environments. To estimate a ``typical ISM value'', we
show the median radiation field and its 10\% to 90\% range in the gas
with density between $1\dim{cm}^{-3}$ and $100\dim{cm}^{-3}$ in the
bottom panel of Fig.\ \ref{fig:gals}. The radiation fields in the ISM
of our model galaxies are significantly above the typical UV intensity
in the Milky Way (which we round off to $J_{\rm
MW}=10^6\dim{photons}/\dim{cm}^2/\dim{s}/\dim{ster}/\dim{eV}$ at
$\lambda=1000{\rm \AA}$, consistent with the measurements of
\citeauthor{h2:d78} \citeyear{h2:d78} and \citeauthor{h2:mmp83}
\citeyear{h2:mmp83}), but in agreement with the typical radiation
fields measured in high-redshift galaxies \citep{hizgal:cppp09}.

\subsection{The Kennicutt-Schmidt relation at high redshifts}

\begin{figure}[t]
\epsscale{1.15}
\plotone{\figname{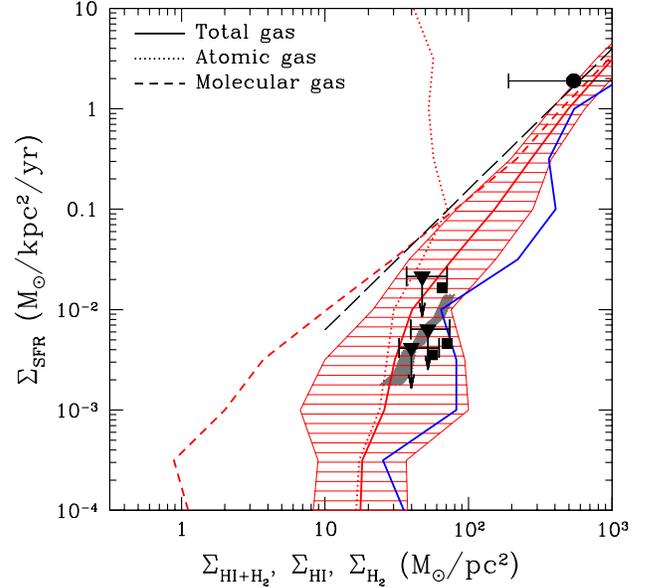}{f3.eps}}
\caption{The KS relation for the simulated galaxies at $z=3$ (red):
  the solid line shows the mean relation, while the hatched band shows
  the RMS scatter around the mean. The total surface density $\Sgas
  \equiv \Shi + \Smol$ takes into account only neutral gas. The dotted
  and short-dashed lines show the mean KS relations for the
  atomic and molecular hydrogen, respectively. A solid blue line shows
  the mean relation only for gas with metallicities below
  $0.1Z_\odot$, which more closely corresponds to the range of
  observed metallicities in the DLAs. The long-dashed line is the best
  fit relation of \citet{sfr:k98a} for $z\approx 0$ galaxies. The gray
  shaded area shows the measurements of the KS relation in LBGs
  (Rafelski et al.\ 2010, in preparation). The solid circle at high gas surface denstity shows measurements 
of SFR and molecular gas surface densities for the lensed $z\approx 3$ LBG cB58 \citep{baker_etal04a}. Solid
  squares and triangles show observational upper limits for the
  $z\approx 3$ DLA systems from \citet{sfr:wc06}, assuming two
  geometrical models for the atomic gas distribution in the DLA
  systems (squares: uniform disks; triangles: isolated clouds). The
  three different points of each type correspond to the three
  representative choices of the smoothing angular scale in Table 1 of
  \citet{sfr:wc06}. Errorbars on the triangles reflect the current
  uncertainty in the high-end slope of the DLA column density
  distribution.} 
\label{fig:sflhz}
\end{figure}

Figure \ref{fig:sflhz} shows the KS relation between star formation
and gas surface densities at $z=3$ for galaxies formed in our
cosmological simulations.  The star formation rate is averaged over
$20\dim{Myr}$, and the gas and SFR surface densities are measured on
the scale of $500\dim{pc}$. We have verified that our results are
robust to changes in the averaging scale (as long as the scale is
$\gtrsim 200$~pc) and the period of time over which the star formation
rate is averaged (for time intervals smaller than $30\dim{Myr}$). The
figure shows that the predicted KS relation for the $z=3$ galaxies is
significantly steeper than the relation for $z=0$ galaxies (the
\citeauthor{sfr:k98a} \citeyear{sfr:k98a} fit to local galaxies is
shown by the dashed line) at $\Sgas \la 100\, \Msun/\dim{pc}^2$ and
its amplitude is an order of magnitude lower than the amplitude of the
\citet{sfr:k98a} fit at $\Sgas \la 50\Msun/\dim{pc}^2$. The
corresponding $\Ssfr-\Smol$ and $\Ssfr-\Shi$ correlations in
Figure~\ref{fig:sflhz} show that the steep $\Ssfr-\Sgas$ relation is
due to the rapidly decreasing ratio of molecular to atomic gas,
$\Smol/\Shi$, at $\Sgas\la 100\, \Msun/\dim{pc}^2$. Likewise, the
surface density at which $\HI$ surface density saturates is $\approx
50\, \Msun/\dim{pc}^2$, considerably larger value than saturation
surface density observed for local galaxies \citep[e.g.,][]{misc:wb02}.

The predicted KS relation for $z\sim 3$ galaxies is consistent with
the recent measurements of star formation densities in Lyman Break
Galaxies by Rafelski et al. (2010, in preparation), which are shown as
a grey band in Fig.\ \ref{fig:sflhz}, and are complemented by the
measurement of $\Ssfr$ and $\Sgas$ for the lensed $z\approx 3$ Lyman
Break Galaxy cB58 \citep{baker_etal04a} at high gas surface density. 

Our simulations are also consistent with the upper limits on star
formation surface density in the $z\approx 3$ DLA systems obtained by
\citet{sfr:wc06}, which are shown in Fig.~\ref{fig:sflhz} with solid
triangles and squares. The three different points of each point type
correspond to the three representative choices of the smoothing
angular scale in Table 1 of \citet{sfr:wc06}.  Solid squares show the
original limits of \citet{sfr:wc06}. However, these limits were
derived assuming a specific model of ``monolithic'' gaseous disks for
DLA systems. This model predicts a specific slope of $-3$ on the high
end of the $\HI$ column density distribution. This value of the slope
is only marginally (at $3\sigma$) consistent with the value of
$-6.4^{+1.1}_{-1.6}$ measured by \citet{prochaska_wolfe09} from the
Data Release 5 (DR5) of the Sloan Digital Sky Survey (SDSS), but is
close to the value of $-3.5$ measured recently by \citet{igm:npls09}.
The two recent measurements indicate that there remains a substantial
uncertainty in the high-$N$ slope of the DLA column density
distribution.

Given the uncertainty in the high-end slope, we also consider an
alternative model for DLAs, in which DLA absorption arises in a
single spherical cloud. The clouds may contain a mixture of atomic and
molecular gas and may have an arbitrary spatial distribution
(including the disk-like), as long as a probability of intersecting
two clouds by a single line of sight is small. This picture may
correspond, for example, to the DLAs arising in the extended HI
envelopes of molecular clouds. In this model any value for the
high-end slope of the column density distribution is allowed. 
These two geometric DLA models bracket the plausible range of possible
geometries and, thus, account for the implicit horizontal uncertainty
of the \citet{sfr:wc06} upper limits.

These limits, recomputed for the single cloud model, are shown in
Fig.\ \ref{fig:sflhz} as downward triangles, with the width of the
errorbars on the triangles reflecting $3\sigma$ errors on the observed
high-end slope from \citet{prochaska_wolfe09}. The error bars also
encompass the \citet{igm:npls09} value and the value of $-3$ for the
monolithic disk model. The error bars of the triangles may serve,
therefore, as a reasonable estimate of the horizonthal uncertainty of
the limits due to the current uncertainty in the high-end DLA column
density distribution. We have converted the \citet{sfr:wc06} upper
limits into linits in the $\Ssfr-\Sgas$ plane by solving their
equations 3 and 6 (using the values from Table 1) for normalization of
the KS relation, $K$, and effective DLA column density, $N$, and
converting $N$ into $\Sgas$.

This conversion itself relies on a subtle assumption that requires
clarification.  DLA lines of sight probe column density averaged on
transverse scale of $\sim 1$~pc, while $\Sgas$ in Fig.~\ref{fig:sflhz}
is averaged on the scale of $500$~pc. The relation between the gas
column densities would of course be direct if the ISM is uniform over
this range of scales, but the ISM density both in real and simulated
galaxies is very far from uniform and exhibits densities ranging by
several orders of magnitude from the average density regions ($n\sim
0.1-1\ \rm cm^{-3}$) to dense molecular clouds ($n\gtrsim 100\ \rm
cm^{-3}$). It may thus seem like there may not be any connection at
all between the column densities measured by the DLAs and average
$\Sgas$, which is a mass-weighted average measured on $\sim$kpc
scale. For example, imagine a simple toy ISM model, in which most of
the mass is in dense, massive clouds with a covering fraction so low
that it is very improbable to intersect them with DLA los.  In this
case, area-weighted surface density can be arbitrarily low, while $\Sgas$ 
averaged on 500~pc scale can be arbitrarily high.

However, there are at least two arguments that correlation between $N$
and $\Sgas$ should be tight even for non-uniform ISM. First, the DLA
column density is an average over tens and hundreds of pc in the los
direction to quasar. Second, average effective $N$ estimated for an
ensemble of DLAs corresponds to the area-weighted average surface
density of ISM in galaxies probed by the DLA lines of sight. For a
realistic probability density distribution (PDF) of ISM the
area-weighted mean surface density should be quite close to the
mass-weighted average.  For example, for the log-normal projected
density PDF, which is a sensible approximation to the PDF of real and
model galaxies \citep[e.g.,][]{elmegreen02}, the area- and
mass-weighted surface densities are directly related \citep[see,
e.g.,][]{ostriker_etal01}. We note, however, that the exact form of
the PDF for real galaxies is yet known and so the factor in converting
$N$ into $\Sgas$ is uncertain. Our conversion and conversion in
\citet{sfr:wc06}, assume a tight one-to-one relation between $N$ and
$\Sgas$.

Figure~\ref{fig:sflhz} shows that the KS relation in our $z=3$ galaxies
is generally consistent with the upper limits of \citet{sfr:wc06},
especially if we recall that metallicities of these galaxies are
somewhat higher than the typical metallicities of DLAs. If we only
consider the gas with $Z \lesssim 0.1Z_\odot$, typical for the DLA
systems, our predictions are fully consistent with the
\citet{sfr:wc06} upper limits.

What is the primary cause of the difference between $z=0$ KS relation
and the $z=3$ relation indicated by observations and our simulations?
The high-redshift galaxies differ from their low redshift counterparts
by their systematically lower metallicities and higher interstellar UV
fluxes.\footnote{There are of course further differences between local
and high redshift galaxies, such as geometry of the gas distribution,
amount of ISM turbulence, etc. These, however, appear to have
relatively minor effect on the KS relation.} The low metallicities are
due to generally lower stellar masses and higher gas fractions of
high-redshift galaxies, which are in the earlier stages of chemical
evolution compared to their local counterparts. The higher UV fluxes
are due to the considerably higher star formation rates, smaller
sizes, and lower abundance of the UV absorbing dust in high redshift
systems.

In principle, both lower metallicity and higher UV flux can affect
star formation in galaxies. Lower metallicity and dust abundance makes
it more difficult to form dense molecular regions in which stars
form\footnote{For example, the transition from atomic to fully
molecular gas occurs at $N_{\rm H} \approx
3\times10^{20}\dim{cm}^{-2}$ in the Milky way, but at $N_{\rm H}
\approx 3\times10^{21}\dim{cm}^{-2}$ in the factor of 5-10 lower
metallicity ISM of the SMC \citep[][this difference is partly due to a
different dust extinction law]{h2:bts03,h2:gstd06}.}
\citep[e.g.,][]{schaye04,sfr:kmt08,sfr:kept09,ng:gtk09,ng:gk10b},
while high UV flux can dissociate such regions
\citep[e.g.,][]{elmegreen93,robertson_kravtsov08,ng:gk10b}.

\begin{figure}[t]
\epsscale{1.15}
\plotone{\figname{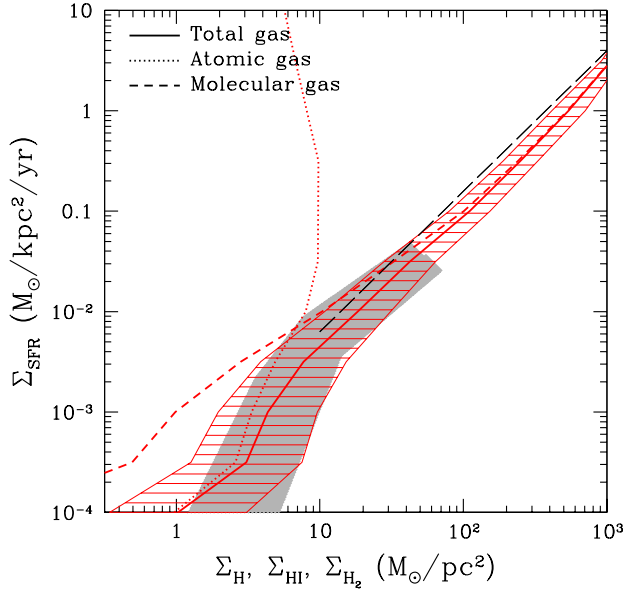}{f4.eps}}
\caption{The KS relation for the simulated galaxies at $z=3$, but with
  dust-to-gas ratio and UV flux fixed to their solar neighborhood
  values. The lines and cross hatched band have the same meaning as in
  Fig.~\ref{fig:sflhz} and the total surface density $\Sigma_{\rm H}$
  takes into account only neutral gas. The gray shaded area shows the
  KS relation for the local dwarf and normal spiral galaxies measured
  by the THINGS project \citep{sfr:blwb08}.}
\label{fig:sflmw}
\end{figure}
To gauge the relative importance of the metallicity and interstellar UV flux
we have rerun our simulation from $z=4$ to $z=2.5$ ($\approx
1\dim{Gyr}$) resetting the dust-to-gas ratio used in the
calculations of H$_2$ abundance and UV field
to fixed multiples of the Milky Way values. We present the results of
such test simulations for a full grid of metallicities and UV
intensities in the companion paper \citep{ng:gk10b}. Here we present
only results for the rerun in which dust-to-gas
abundance and UV flux were set to their Milky Way values. 
Figure~\ref{fig:sflmw} shows the KS relation for this test re-simulation at
$z=3$. As before, the star formation rate in the simulations is
averaged over $20\dim{Myr}$ and the gas and SFR surface densities are
averaged on the scale of $500$~pc. The figure shows that in the case
of the MW metallicity and UV flux the predicted KS relation is
consistent with the observed relation for the $z=0$ galaxies
\citep{sfr:k98a,sfr:blwb08}. In particular, the predicted KS relation
matches the approximately linear relation relation $\Ssfr \propto
\Sgas^{1.0}$ at $\Sgas\ga 10 \Msun/\dim{pc}^2$ (with indication of
steepening to $\Ssfr \propto \Sgas^{1.4\div 1.6}$ at $\Sgas\ga 10^2
\Msun/\dim{pc}^2$) and the significant steepening of the relation at
$\Sgas\la 10\ \rm M_{\odot}\,pc^{-2}$. Simulations also reproduce the
observed scatter around the mean relation reasonably well. Finally,
note that the atomic hydrogen in this case saturates at surface
density of $\Shi\approx 10\ \rm M_{\odot}\,pc^{-2}$, which is
consistent with observations of nearby galaxies
\citep[e.g.,][]{misc:wb02,sfr:blwb08} and a factor of five smaller than
the corresponding value for the self-consistent metallicities and UV
fluxes of the simulated $z=3$ galaxies in Fig.\ \ref{fig:sflhz}.

The fact that the local KS relation is reproduced in simulations of
high redshift disks with higher metallicity and lower UV flux indicates
that the differences in the ISM structure between high and low
redshift galaxies play a relatively minor role in shaping this
relation.  Although the density PDF in gaseous disks can be important
in shaping the slope of the KS relation
\citep{kravtsov03,tassis07,robertson_kravtsov08}, our test shows that
the main factors behind the difference between $z=0$ and $z=3$ KS
relations are gas metallicity and UV flux in the ISM. Furthermore,
additional experiments, presented in the companion paper
\citep{ng:gk10b}, show that the difference in metallicity is the
primary reason behind the differences in the KS relations shown in
figures~\ref{fig:sflhz} and \ref{fig:sflmw}. 

The UV flux does affect star formation in the individual patches of
the ISM significantly, but it also affects the surface density of
atomic and molecular gas. The changes of $\Ssfr$ and $\Sgas = \Shi +
\Smol$ affected by the change in the UV flux are such that a given
patch of ISM is moving approximately along the KS relation for a varying FUV flux. The
average KS relation is therefore less sensitive to the changes in UV
flux than to the changes in the metallicity of the same magnitude,
even though star formation in individual ISM regions and galaxies is
strongly affected. The change of metallicities, however,
affects $\Ssfr$ without affecting the surface density of neutral gas
significantly, thereby changing the mean KS relation. The effect of the UV flux 
on the KS relation does increase for
lower metallicity galaxies, as we discuss in the companion paper.

The conclusion from our experiments is therefore that the main reason
the KS relation of high redshift galaxies is significantly different
from their low-redshift counterparts is their lower metallicity and
dust abundance, while the higher FUV flux and differences in density
PDFs are secondary factors.

\section{Discussion and Conclusions}
\label{sec:discussion}

In this study we use recently developed recipe for star formation in
simulations of galaxy formation based on the local abundance of
molecular hydrogen (eq.~\ref{eq:sfr}) to examine the Kennicutt-Schmidt
relation expected in high redshift systems. The $\H2$ abundance is
tracked self-consistently in the simulation using a phenomenological
model, which includes effects of dissociating UV radiation, as well as
self-shielding and shielding of $\H2$ by dust \citep[see][for details
of star formation recipe and $\H2$ model]{ng:gtk09,ng:gk10b}. 

Our main result, presented in Fig.~\ref{fig:sflhz}, is that the KS
relation in the low-metallicity and high UV flux environments of
$z\sim 3$ is substantially steeper and has a lower amplitude than that
of the $z=0$ galaxies at $\Sgas\la 100\rm\ M_{\odot}\,pc^{-2}$. As
discussed in the previous section, the main reason for the difference
is lower metallicity and dust-to-gas ratio of high redshift galaxies. While our tests show that the UV flux
does affect the star formation drastically, it simultaneously
affects surface density of neutral gas in such a way that the shape and normalization of the
KS relation are largely unaffected.

\begin{figure}[t]
\epsscale{1.1}
\plotone{\figname{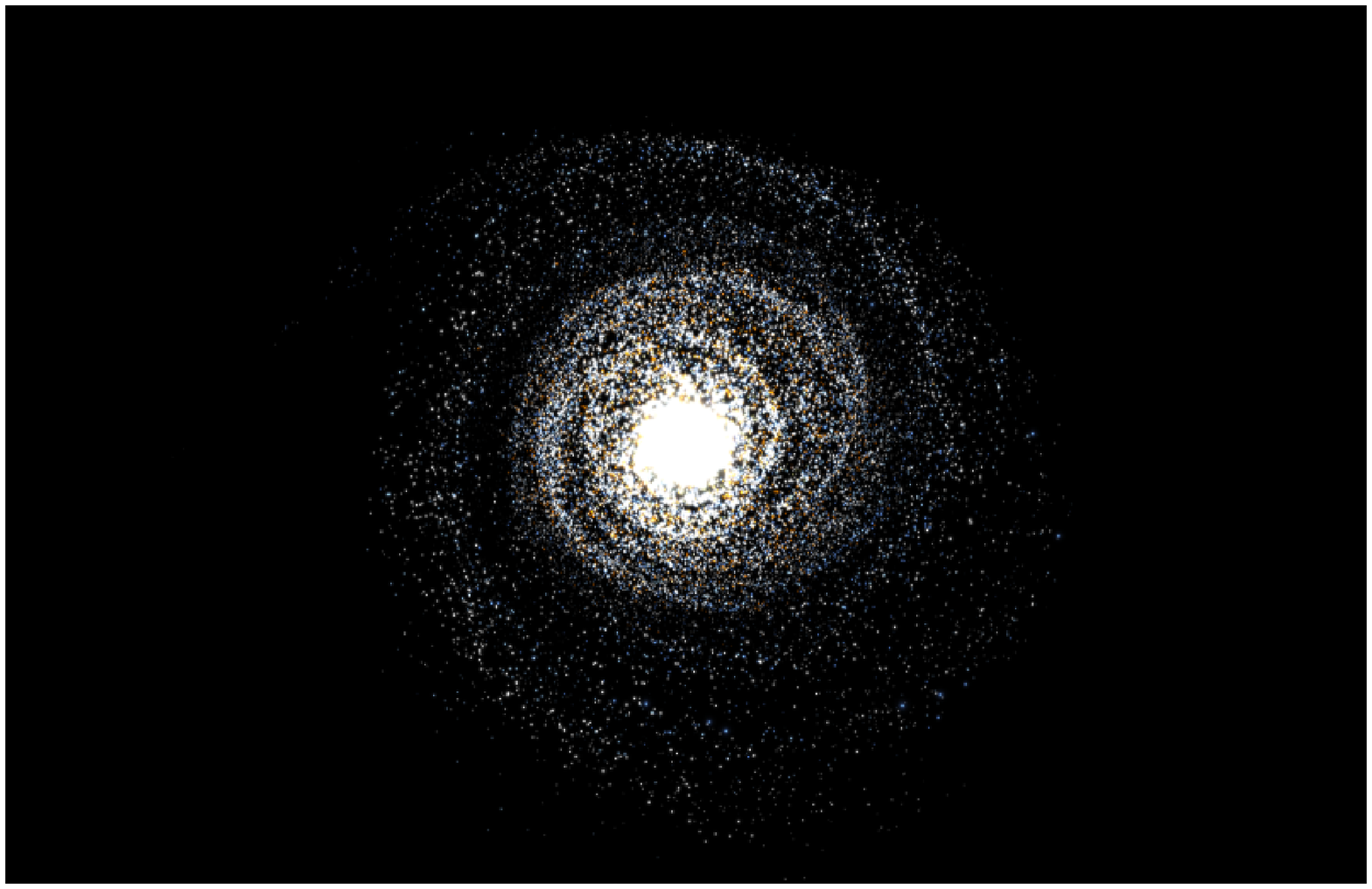}{f5a.eps}}
\plotone{\figname{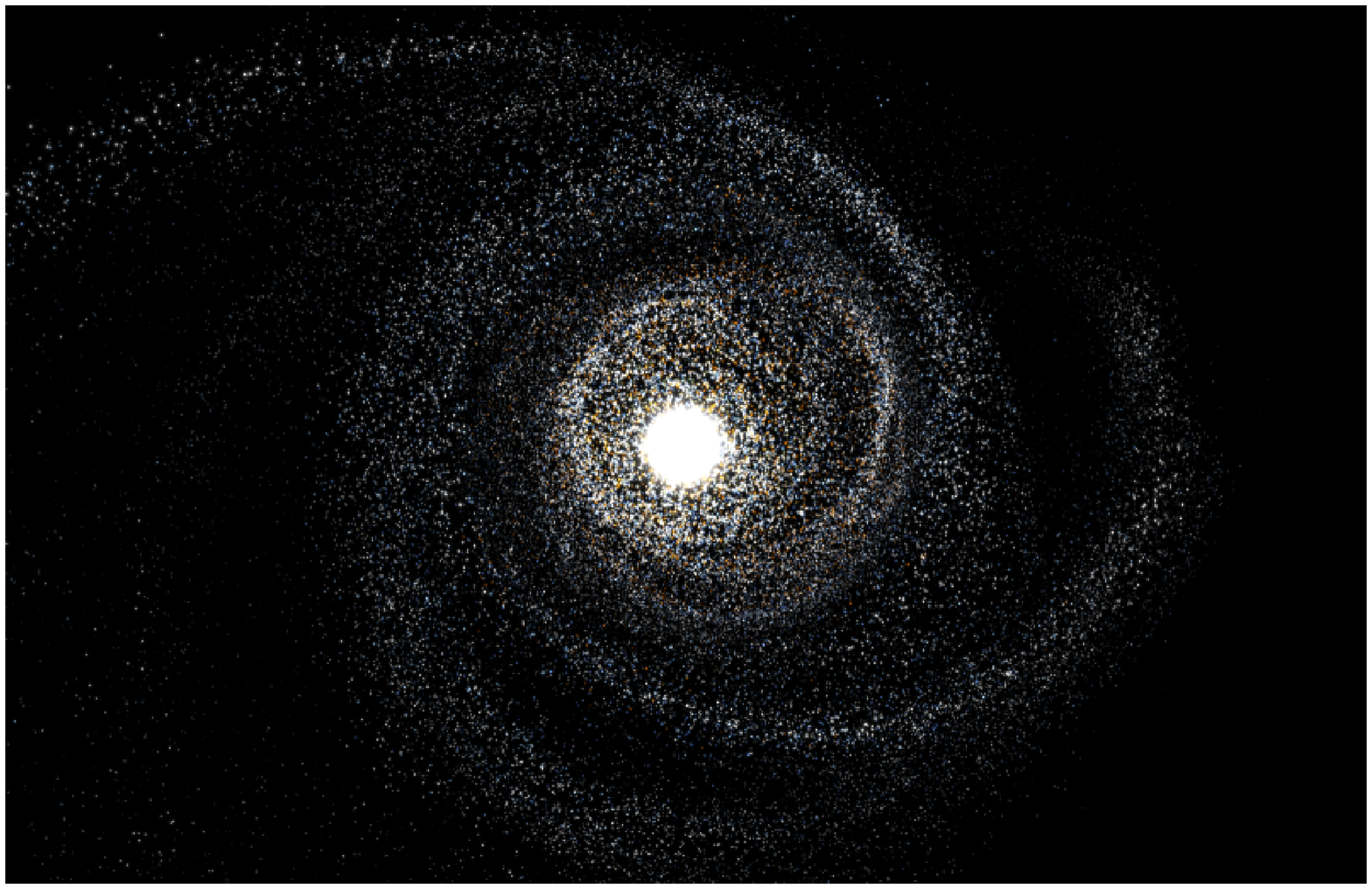}{f5b.eps}}
\caption{The face-on view of the disk of young stars of the most
massive galaxy in the simulation at $z=2.5$. Only stars younger than
$50\dim{Myr}$ are shown; the colors are indicating the B-V color
(bright white corresponding to the bluest, youngest stars). The
vertical size of each panel is identical and is approximately equal to
10 physical kpc. The top panel shows result of the self-consistent
simulation (i.e., the simulation for which the KS relations shown in
Fig.~\ref{fig:sflhz}), while the bottom panel shows result of the test
run, in which dust abundance and FUV flux were set to their Milky Way
values at $z=4$ and the simulation was run with these fixed values to
$z=2.5$ (KS relation for this run is shown in
Fig.~\ref{fig:sflmw}). Note that star formation in the self-consistent
case is concentrated towards higher gas density regions, which results
in more compact distribution of stars in the disk and fewer young
stars in the inter-arm regions, as shown in Fig.~\ref{fig:sbpro}.}
\label{fig:starviz}
\end{figure}

\begin{figure}[t]
\epsscale{1.3}
\plotone{\figname{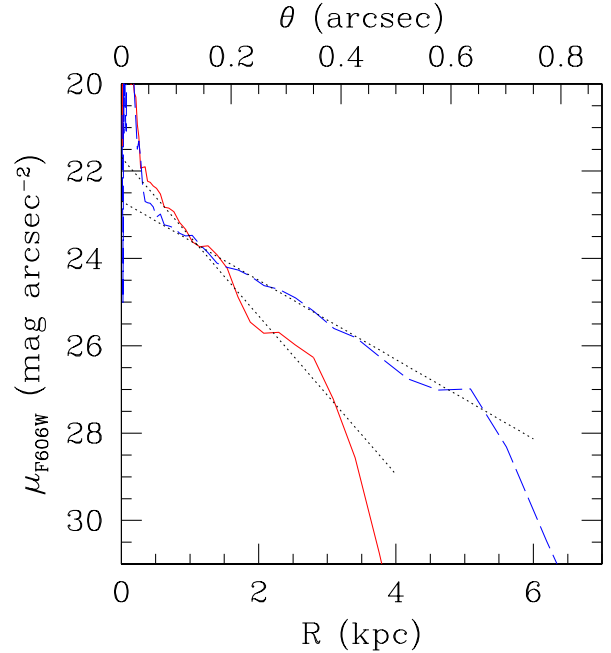}{f6.eps}}
\caption{Surface brightness profiles of the most massive galaxy at
$z=2.5$ in the self-consistent simulation (i.e., the simulation for
which the KS relations shown in Fig.~\ref{fig:sflhz}; {\it red solid
line}), and in the test run in which dust abundance and FUV flux were
set to their Milky Way values at $z=4$ and the simulation was run with
these fixed values to $z=2.5$ (KS relation for this run is shown in
Fig.~\ref{fig:sflmw}; {\it dashed blue line}). The bottom $x$-axis is
in physical kpc, while the top axis shows the corresponding angular
scale in arcseconds. The surface brightness, $\mu_{\rm F606W}$, was
computed using the Flexible SPS code of Conroy \& Gunn (2009) in AB
magnitudes for the ACS F606W filter by appropriately redshifting the
spectrum of emission to $z=2.5$ (the light in this filter is thus
dominated by the rest-frame FUV radiation) and takes into account
cosmological $(1+z)^4$ dimming, but does not take into account any
dust absorption. For comparison, the dotted lines show exponential
disk profiles with $R_e=1$ and $2$~kpc.}
\label{fig:sbpro}
\end{figure}

Our predicted KS relation for $z\sim 3$ galaxies agrees well with a
recent estimate of $\Ssfr$ in the Lyman Break Galaxies by Rafelski et
al. (2009, in preparation) at low gas surface densities and
measurements of SFR and gas surface density in LBG cB58 at high
surface densities \citep{baker_etal04a}.  Our results are also
consistent with the upper limits on star formation surface density
derived for the subset of $z\sim 3$ DLA systems at gas surface
densities of $\Sgas\sim 20-50\rm\ M_{\odot}\,pc^{-2}$
\citep{sfr:wc06}. These limits are lower by a factor of ten compared
to the $\Ssfr$ for the local galaxies at the same gas surface
densities \citep{sfr:k98a} similar to our simulation results. We note,
however, that the typical metallicity of the gas shown in
Fig.~\ref{fig:gals} is a few times higher than the typical metallicity
of the DLA systems
\citep[e.g.,][]{prochaska_etal03,fynbo_etal08}. Given that effect of
decreasing metallicity on the KS relation is to steepen it at low
surface densities and lower its amplitude, we can expect that for
systems of lower metallicity than those we have considered in this
study the KS relation should have even steeper slope and lower
amplitude, as is shown by the solid blue line in Fig.~\ref{fig:sflhz}.

A qualitatively similar trend of the KS relation with metallicity was
recently predicted by \citet{krumholz_etal09} using a model of atomic
to molecular transition in molecular complexes based on the
\citet{wolfire_etal03} semi-analytic model of atomic ISM \citep[see also][]{mckee_krumholz09}. Their model
predicts significant steepening of the KS relation below gas surface
density of $\Sigma_{\rm H}\lesssim 10/cZ\,\rm M_{\odot}\,pc^{-2}$,
where $Z$ is the metallicity of the gas in the units of $Z_{\odot}$
and $c$ is the ISM clumping factor. The latter reflects the difference between
the surface density of the ISM averaged on some scale $>100$~pc and
the surface density of individual giant molecular complexes on the
scale of $100$~pc. For averaging scales of $500$~pc used in our
calculations the clumping factor should be $c\sim 2-5$
\citep{krumholz_etal09}.  The model thus predicts the steepening of
the KS relation at $\Sgas\approx 2-5\,\rm M_{\odot}\,pc^{-2}$
for solar metallicities and $\Sgas\approx 6-15\,\rm
M_{\odot}\,pc^{-2}$ for $Z\approx 0.2-0.5Z_{\odot}$ characteristic for
our simulated galaxies. These values are considerably lower than the
transition surface density of $\Sgas\approx 50-100\,\rm
M_{\odot}\,pc^{-2}$ we derive for our simulated galaxies. It is not
entirely clear what the source of this difference is. We note,
however, that the mass fraction of molecular gas for a given
large-scale surface density $\Sgas$ depends on the shape of
the density PDF of the ISM, which may well depend on the gas surface
density \citep[e.g.,][]{kravtsov03}. This dependence may not be
captured by a single clumping factor independent of $\Sgas$.

The strong trend of the KS relation with metallicity has several
interesting implications for theoretical scenarios of galaxy formation
and interpretation of observations of high redshift galaxies\footnote{For example, significantly different KS relation for high redshift 
galaxies may strongly bias gas mass estimates that use $z=0$ calibration
of this relation for systems of metallicity significantly lower than
solar \citep[e.g.,][]{erb_etal06,manucci_etal09}.}. The trend
implies, for example, that star formation in lower-metallicity
high redshift galaxies should be more centrally concentrated compared to
their local counterparts. In particular, the regions of high redshift disks
with $\Sgas\la 100\rm\ M_{\odot}\,pc^{-2}$ should be considerably
dimmer than expected from the $z=0$ KS relation. 

This trend is illustrated in Figures~\ref{fig:starviz} and
\ref{fig:sbpro}, which show distribution of young stars and rest-frame
FUV surface brightness profiles (computed using the Flexible SPS code
of \citeauthor{conroy_gunn09} \citeyear{conroy_gunn09}) in the most
massive galaxy in our self-consistent simulations compared to the test
run, in which dust-to-gas ratio and UV flux were fixed to their MW
values between $z=4$ and $z=2.5$. The galaxies in this test run
exhibit KS relation consistent with the local measurements (see
Fig.~\ref{fig:sflmw}). The star formation in the self-consistent run
is concentrated towards the higher gas surface density central
regions.

Figure~\ref{fig:sbpro} shows that the apparent size of the
simulated galaxy sensitively depends on the physics of molecular gas
and the form of the KS relation. In both self-consistent and test
runs, the rest-frame UV emission of the simulated galaxy is well
described by the extended exponential disk profile (dotted lines). The
effective half-light radii of the disks, however, are different by a factor of
two: $R_e=1$~kpc and $2$~kpc in the self-consistent and test runs,
respectively. This illustrates that to predict the observed sizes of galaxies,
it is not sufficient to model dynamics of gas condensation and
acquisition of angular momentum correctly. Correct modeling of star
formation in gas of different densities is equally
important. 

Incidentally, the search for extended low surface brightness emission
in the HUDF carried out by \citet{sfr:wc06} did not identify any objects 
that had isophotal radii $R_{28.4}>4$~kpc corresponding
to $\mu_{\rm F606W}=28.4$. Figure~\ref{fig:sbpro} shows that surface
brightness profile of the galaxy in the self-consistent simulations is
consistent with the observed paucity of such extended objects in the
HUDF, while the profile of the galaxy in the test run is not consistent with
that search. This is another illustration of the fact that the local
KS relation would predict existence of extended low surface brightness
disks at $z\sim 3$ that are not observed in the HUDF, as emphasized by
\citet{sfr:wc06}.

The emerging picture based on observational constraints
\citep{sfr:wc06,baker_etal04b,wolfe_etal04,wolfe_etal08} and our simulations is therefore that DLA systems
correspond to the lower metallicity low-mass halos in which there is
little or no ongoing star formation or to the outskirts of low
metallicity disks surrounding centrally-concentrated, actively star
forming Lyman Break Galaxies.  The latter picture is qualitatively similar to the
low-metallicity $z=0$ dwarf galaxies, which often have a very compact
stellar system embedded in an extended gas-rich disk
\citep[e.g.,][]{meurer_etal96,begum_etal06}.  In this picture the
neutral atomic gas probed in absorption is largely inert to star
formation, but actively star forming regions providing metals and UV
heating may be located in the center or in spiral arms a few
kiloparsecs away within the same system.  

The long gas consumption time scales at $\Sgas\la 100\rm\
M_{\odot}\,pc^{-2}$ would allow newly accreted gas to accumulate at
these relatively low gas surface densities, instead of being consumed
on a $\approx 1-2$~Gyr time scale.  This can result in a more extended star formation history if the host
galaxy evolves quiescently or provide gas reservoir for a starburst if
the host object undergoes a major merger. It is also possible that
accumulating gas in the disk undergoes a starburst as it is enriched
by in situ star formation and its surface density increases due to
accretion of new gas, thereby creating conditions for gravitational
instabilities and widespread star formation
\citep[e.g.,][]{bournaud_etal07,agertz_etal09,dekel_etal09}.  The
overall effect of considerably suppressed star formation at $\Sgas\la
100\rm\ M_{\odot}\,pc^{-2}$ is to preserve more gas in smaller mass
objects for star formation in galaxies forming the bulk of their stars
at lower ($z\la 2$) redshifts.

Another potentially very important implication of the more centrally
concentrated star formation is that the stars formed at high-redshifts
are confined to the high surface density regions and are therefore
more resistant against dynamical heating in mergers. The outer {\it
gaseous\/} disks are resilient to major mergers
\citep[e.g..][]{robertson_etal04,robertson_etal06,springel_hernquist05}
and would help maintain nascent stellar disks dynamically cold during
minor mergers \citep{moster_etal09b}, thereby facilitating formation
of thin stellar disks at later epochs and helping explain the
prevalence of such disks in the local universe.

\begin{figure}[t]
\epsscale{1.15}
\plotone{\figname{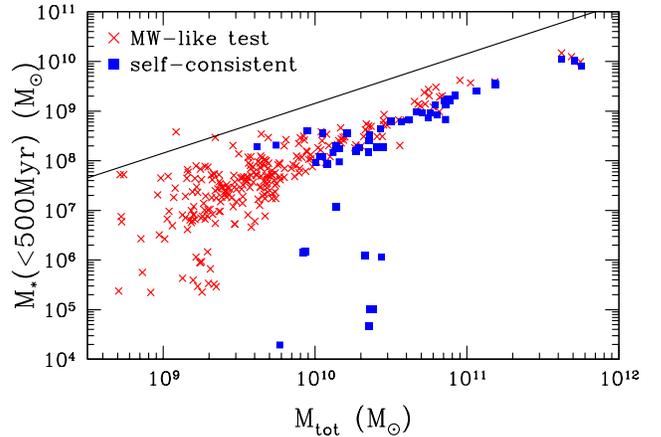}{f7.eps}}
\caption{The stellar mass formed in the last $500\dim{Myr}$ in the simulated galaxies as a
  function of the total virial mass of their host halo for the full cosmological run (red
  circles) and the test run (blue squares). To get a representative
  sample, we show three simulation outputs together, at $a=0.24$,
  $0.26$, and $0.28$. These outputs are separated by about
  $250\dim{Myr}$ and, thus, only partially correlated.}
\label{fig:msmt}
\end{figure}

All these effects should be particularly salient in dwarf galaxies
($M_{\rm tot}\la 10^{11}\rm\ M_{\odot}$), which have the lowest
metallicities and surface gas densities and where suppression of star
formation could thus expected to be particularly strong. We do indeed
see a progressively lower efficiency of star formation in halos of
lower mass (Fig.~\ref{fig:masses}), and this decrease is not due to
supernova feedback (as demonstrated by a comparison with test
simulations in which the supernova feedback is artificially
disabled). The fact that stellar mass is significantly decreased for a
given $M_{\rm tot}$ in our star formation model compared to the
standard star formation prescription is demonstrated in
Figure~\ref{fig:msmt}.  This figure shows masses of stars in simulated
galaxies younger than 500 Myrs versus the total virial mass of their
host halos in the full self-consistent simulation and in the test run
in which dust-to-gas ration and FUV field were fixed to their Milky
Way values. The latter test is equivalent to simply using the fixed
density threshold for star forming gas, as is usually assumed in
galaxy formation simulations.

The figure shows that self-consistent simulation with
our H$_2$-based star formation model produces systematically smaller
stellar masses in simulated galaxies compared to the test run. The
difference is only $\approx 20-30\%$ for the largest galaxies, but
increases to a factor of two and more for smaller
galaxies. Remarkably, for halo masses of $M_{\rm tot}\lesssim
10^{10}\,\rm M_{\odot}$ the star formation is completely suppressed in
the full simulation, while in the test simulation star formation is unabated 
down to smallest resolved halo masses, well
below $10^{10}\,\rm M_{\odot}$. This shows that low dust abundance and
high FUV field can play a major role in suppressing star formation in
low metallicity, low mass halos -- the effect that can help explaining
the difference between predicted steep small-mass slope of the halo
mass function and the shallow observed faint-end slope of the
luminosity function. Note that this suppression is achieved without efficient supernova feedback. 

The results we presented in this paper thus indicate that other
processes in the ISM in addition (or perhaps in lieu) of the standard
supernova feedback may be shaping the evolution of galaxy population
and setting the observed disk sizes. Investigating these processes is
an interesting avenue for future research. Theoretical investigations
in this direction would be greatly aided by the observations of
molecular gas in high-redshift star forming galaxies provided by the
CARMA and ALMA observatories in the near future.

\acknowledgements 

We thank Hsiao-Wen Chen, Marc Rafelski, and Art Wolfe for important
comments and corrections on the earlier versions of this paper. We are
also indebted to Marc and Art for the permission to show their yet
unpublished measurements in our Fig.\ \ref{fig:sflhz} and for their
help in translating \citet{sfr:wc06} limits into points shown in Fig.\
\ref{fig:sflhz}. We would like to thank Charlie Conroy for making his
FSPS code available and assistance with the SPS calculations. AK would
like to thank organizers of the ``SFR@50'' conference in July 2009 and
owners of the Abbazia di Spineto for the great meeting and wonderful
atmosphere for scientific exchange, from which this paper has
benefited greatly.  This work was supported in part by the DOE at
Fermilab, by the NSF grants AST-0507596 and AST-0708154, and by the
Kavli Institute for Cosmological Physics at the University of Chicago
through the NSF grant PHY-0551142 and an endowment from the Kavli
Foundation. The simulations used in this work have been performed on
the Joint Fermilab - KICP Supercomputing Cluster, supported by grants
from Fermilab, Kavli Institute for Cosmological Physics, and the
University of Chicago.  This work made extensive use of the NASA
Astrophysics Data System and {\tt arXiv.org} preprint server.

\bibliographystyle{apj}
\bibliography{ak,ng-bibs/sfr,ng-bibs/self,ng-bibs/sims,ng-bibs/h2,ng-bibs/igm,ng-bibs/hizgal,ng-bibs/misc,ng-bibs/dsh,ng-bibs/cosmo}

\begin{thebibliography}{93}
\expandafter\ifx\csname natexlab\endcsname\relax\def\natexlab#1{#1}\fi

\bibitem[{{Agertz} {et~al.}(2009){Agertz}, {Teyssier}, \&
  {Moore}}]{agertz_etal09}
{Agertz}, O., {Teyssier}, R., \& {Moore}, B. 2009, \mnras, 397, L64

\bibitem[{{Baker} {et~al.}(2004{\natexlab{a}}){Baker}, {Tacconi}, {Genzel},
  {Lehnert}, \& {Lutz}}]{baker_etal04a}
{Baker}, A.~J., {Tacconi}, L.~J., {Genzel}, R., {Lehnert}, M.~D., \& {Lutz}, D.
  2004{\natexlab{a}}, \apj, 604, 125

\bibitem[{{Baker} {et~al.}(2004{\natexlab{b}}){Baker}, {Tacconi}, {Genzel},
  {Lutz}, \& {Lehnert}}]{baker_etal04b}
{Baker}, A.~J., {Tacconi}, L.~J., {Genzel}, R., {Lutz}, D., \& {Lehnert}, M.~D.
  2004{\natexlab{b}}, \apjl, 613, L113

\bibitem[{{Begum} {et~al.}(2006){Begum}, {Chengalur}, {Karachentsev}, {Kaisin},
  \& {Sharina}}]{begum_etal06}
{Begum}, A., {Chengalur}, J.~N., {Karachentsev}, I.~D., {Kaisin}, S.~S., \&
  {Sharina}, M.~E. 2006, \mnras, 365, 1220

\bibitem[{{Bigiel} {et~al.}(2008){Bigiel}, {Leroy}, {Walter}, {Brinks}, {de
  Blok}, {Madore}, \& {Thornley}}]{sfr:blwb08}
{Bigiel}, F., {Leroy}, A., {Walter}, F., {Brinks}, E., {de Blok}, W.~J.~G.,
  {Madore}, B., \& {Thornley}, M.~D. 2008, \aj, 136, 2846

\bibitem[{{Birnboim} \& {Dekel}(2003)}]{birnboim_dekel03}
{Birnboim}, Y. \& {Dekel}, A. 2003, \mnras, 345, 349

\bibitem[{{Blitz} \& {Rosolowsky}(2006)}]{blitz_rosolowsky06}
{Blitz}, L. \& {Rosolowsky}, E. 2006, \apj, 650, 933

\bibitem[{{Blumenthal} {et~al.}(1984){Blumenthal}, {Faber}, {Primack}, \&
  {Rees}}]{blumenthal_etal84}
{Blumenthal}, G.~R., {Faber}, S.~M., {Primack}, J.~R., \& {Rees}, M.~J. 1984,
  \nat, 311, 517

\bibitem[{{Boissier} {et~al.}(2007){Boissier}, {Gil de Paz}, {Boselli},
  {Madore}, {Buat}, {Cortese}, {Burgarella}, {Mu{\~n}oz-Mateos}, {Barlow},
  {Forster}, {Friedman}, {Martin}, {Morrissey}, {Neff}, {Schiminovich},
  {Seibert}, {Small}, {Wyder}, {Bianchi}, {Donas}, {Heckman}, {Lee},
  {Milliard}, {Rich}, {Szalay}, {Welsh}, \& {Yi}}]{boissier_etal07}
{Boissier}, S., {Gil de Paz}, A., {Boselli}, A., {Madore}, B.~F., {Buat}, V.,
  {Cortese}, L., {Burgarella}, D., {Mu{\~n}oz-Mateos}, J.~C., {Barlow}, T.~A.,
  {Forster}, K., {Friedman}, P.~G., {Martin}, D.~C., {Morrissey}, P., {Neff},
  S.~G., {Schiminovich}, D., {Seibert}, M., {Small}, T., {Wyder}, T.~K.,
  {Bianchi}, L., {Donas}, J., {Heckman}, T.~M., {Lee}, Y.-W., {Milliard}, B.,
  {Rich}, R.~M., {Szalay}, A.~S., {Welsh}, B.~Y., \& {Yi}, S.~K. 2007, \apjs,
  173, 524

\bibitem[{{Bolatto}(2009)}]{bolatto_etal09}
{Bolatto}, A. 2009, in preparation

\bibitem[{{Bouch{\'e}} {et~al.}(2007){Bouch{\'e}}, {Cresci}, {Davies},
  {Eisenhauer}, {F{\"o}rster Schreiber}, {Genzel}, {Gillessen}, {Lehnert},
  {Lutz}, {Nesvadba}, {Shapiro}, {Sternberg}, {Tacconi}, {Verma}, {Cimatti},
  {Daddi}, {Renzini}, {Erb}, {Shapley}, \& {Steidel}}]{sfr:b07}
{Bouch{\'e}}, N., {Cresci}, G., {Davies}, R., {Eisenhauer}, F., {F{\"o}rster
  Schreiber}, N.~M., {Genzel}, R., {Gillessen}, S., {Lehnert}, M., {Lutz}, D.,
  {Nesvadba}, N., {Shapiro}, K.~L., {Sternberg}, A., {Tacconi}, L.~J., {Verma},
  A., {Cimatti}, A., {Daddi}, E., {Renzini}, A., {Erb}, D.~K., {Shapley}, A.,
  \& {Steidel}, C.~C. 2007, \apj, 671, 303

\bibitem[{{Bournaud} {et~al.}(2007){Bournaud}, {Elmegreen}, \&
  {Elmegreen}}]{bournaud_etal07}
{Bournaud}, F., {Elmegreen}, B.~G., \& {Elmegreen}, D.~M. 2007, \apj, 670, 237

\bibitem[{{Browning} {et~al.}(2003){Browning}, {Tumlinson}, \&
  {Shull}}]{h2:bts03}
{Browning}, M.~K., {Tumlinson}, J., \& {Shull}, J.~M. 2003, \apj, 582, 810

\bibitem[{{Chen} {et~al.}(2009){Chen}, {Perley}, {Pollack}, {Prochaska},
  {Bloom}, {Dessauges-Zavadsky}, {Pettini}, {Lopez}, {Dall'aglio}, \&
  {Becker}}]{hizgal:cppp09}
{Chen}, H.-W., {Perley}, D.~A., {Pollack}, L.~K., {Prochaska}, J.~X., {Bloom},
  J.~S., {Dessauges-Zavadsky}, M., {Pettini}, M., {Lopez}, S., {Dall'aglio},
  A., \& {Becker}, G.~D. 2009, \apj, 691, 152

\bibitem[{{Conroy} \& {Gunn}(2009)}]{conroy_gunn09}
{Conroy}, C. \& {Gunn}, J.~E. 2009, ArXiv e-prints

\bibitem[{{Conroy} \& {Wechsler}(2009)}]{conroy_wechsler09}
{Conroy}, C. \& {Wechsler}, R.~H. 2009, \apj, 696, 620

\bibitem[{{Conroy} {et~al.}(2006){Conroy}, {Wechsler}, \&
  {Kravtsov}}]{conroy_etal06}
{Conroy}, C., {Wechsler}, R.~H., \& {Kravtsov}, A.~V. 2006, \apj, 647, 201

\bibitem[{{Dekel} {et~al.}(2009){Dekel}, {Sari}, \& {Ceverino}}]{dekel_etal09}
{Dekel}, A., {Sari}, R., \& {Ceverino}, D. 2009, \apj, 703, 785

\bibitem[{{Draine}(1978)}]{h2:d78}
{Draine}, B.~T. 1978, \apjs, 36, 595

\bibitem[{{Dunkley} {et~al.}(2009){Dunkley}, {Komatsu}, {Nolta}, \& {et
  al.}}]{dunkley_etal09}
{Dunkley}, J., {Komatsu}, E., {Nolta}, M.~R., \& {et al.} 2009, \apjs, 180, 306

\bibitem[{{Elmegreen}(1993)}]{elmegreen93}
{Elmegreen}, B.~G. 1993, \apj, 411, 170

\bibitem[{{Elmegreen}(2002)}]{elmegreen02}
---. 2002, \apj, 577, 206

\bibitem[{{Elmegreen} \& {Parravano}(1994)}]{elmegreen_parravano94}
{Elmegreen}, B.~G. \& {Parravano}, A. 1994, \apjl, 435, L121+

\bibitem[{{Erb} {et~al.}(2006{\natexlab{a}}){Erb}, {Steidel}, {Shapley},
  {Pettini}, {Reddy}, \& {Adelberger}}]{hizgal:essp06}
{Erb}, D.~K., {Steidel}, C.~C., {Shapley}, A.~E., {Pettini}, M., {Reddy},
  N.~A., \& {Adelberger}, K.~L. 2006{\natexlab{a}}, \apj, 646, 107

\bibitem[{{Erb} {et~al.}(2006{\natexlab{b}}){Erb}, {Steidel}, {Shapley},
  {Pettini}, {Reddy}, \& {Adelberger}}]{erb_etal06}
---. 2006{\natexlab{b}}, \apj, 646, 107

\bibitem[{{Fox} {et~al.}(2009){Fox}, {Prochaska}, {Ledoux}, {Petitjean},
  {Wolfe}, \& {Srianand}}]{igm:fplp09}
{Fox}, A.~J., {Prochaska}, J.~X., {Ledoux}, C., {Petitjean}, P., {Wolfe},
  A.~M., \& {Srianand}, R. 2009, \aap, 503, 731

\bibitem[{{Fynbo} {et~al.}(2008){Fynbo}, {Prochaska}, {Sommer-Larsen},
  {Dessauges-Zavadsky}, \& {M{\o}ller}}]{fynbo_etal08}
{Fynbo}, J.~P.~U., {Prochaska}, J.~X., {Sommer-Larsen}, J.,
  {Dessauges-Zavadsky}, M., \& {M{\o}ller}, P. 2008, \apj, 683, 321

\bibitem[{{Gillmon} {et~al.}(2006){Gillmon}, {Shull}, {Tumlinson}, \&
  {Danforth}}]{h2:gstd06}
{Gillmon}, K., {Shull}, J.~M., {Tumlinson}, J., \& {Danforth}, C. 2006, \apj,
  636, 891

\bibitem[{{Gnedin} \& {Abel}(2001)}]{ng:ga01}
{Gnedin}, N.~Y. \& {Abel}, T. 2001, New Astronomy, 6, 437

\bibitem[{{Gnedin} \& {Kravtsov}(2010)}]{ng:gk10b}
{Gnedin}, N.~Y. \& {Kravtsov}, A.~V. 2010, in preparation

\bibitem[{{Gnedin} {et~al.}(2009){Gnedin}, {Tassis}, \& {Kravtsov}}]{ng:gtk09}
{Gnedin}, N.~Y., {Tassis}, K., \& {Kravtsov}, A.~V. 2009, \apj, 697, 55

\bibitem[{{Heyer} {et~al.}(2004){Heyer}, {Corbelli}, {Schneider}, \&
  {Young}}]{heyer_etal04}
{Heyer}, M.~H., {Corbelli}, E., {Schneider}, S.~E., \& {Young}, J.~S. 2004,
  \apj, 602, 723

\bibitem[{{Kennicutt}(1989)}]{sfr:k89}
{Kennicutt}, Jr., R.~C. 1989, \apj, 344, 685

\bibitem[{{Kennicutt}(1998)}]{sfr:k98a}
---. 1998, \apj, 498, 541

\bibitem[{{Kennicutt} {et~al.}(2007){Kennicutt}, {Calzetti}, {Walter}, {Helou},
  {Hollenbach}, {Armus}, {Bendo}, {Dale}, {Draine}, {Engelbracht}, {Gordon},
  {Prescott}, {Regan}, {Thornley}, {Bot}, {Brinks}, {de Blok}, {de Mello},
  {Meyer}, {Moustakas}, {Murphy}, {Sheth}, \& {Smith}}]{kennicutt_etal07}
{Kennicutt}, Jr., R.~C., {Calzetti}, D., {Walter}, F., {Helou}, G.,
  {Hollenbach}, D.~J., {Armus}, L., {Bendo}, G., {Dale}, D.~A., {Draine},
  B.~T., {Engelbracht}, C.~W., {Gordon}, K.~D., {Prescott}, M.~K.~M., {Regan},
  M.~W., {Thornley}, M.~D., {Bot}, C., {Brinks}, E., {de Blok}, E., {de Mello},
  D., {Meyer}, M., {Moustakas}, J., {Murphy}, E.~J., {Sheth}, K., \& {Smith},
  J.~D.~T. 2007, \apj, 671, 333

\bibitem[{{Kere{\v s}} {et~al.}(2005){Kere{\v s}}, {Katz}, {Weinberg}, \&
  {Dav{\'e}}}]{keres_etal05}
{Kere{\v s}}, D., {Katz}, N., {Weinberg}, D.~H., \& {Dav{\'e}}, R. 2005,
  \mnras, 363, 2

\bibitem[{{Kravtsov}(1999)}]{kravtsov99}
{Kravtsov}, A.~V. 1999, PhD thesis, AA(NEW MEXICO STATE UNIVERSITY)

\bibitem[{{Kravtsov}(2003)}]{kravtsov03}
---. 2003, \apjl, 590, L1

\bibitem[{{Kravtsov}(2006)}]{kravtsov06}
---. 2006, proceedings of the XLIst Rencontres de Moriond (arXiv/0607463)

\bibitem[{{Kravtsov} {et~al.}(2002){Kravtsov}, {Klypin}, \&
  {Hoffman}}]{kravtsov_etal02}
{Kravtsov}, A.~V., {Klypin}, A., \& {Hoffman}, Y. 2002, \apj, 571, 563

\bibitem[{{Krumholz} {et~al.}(2009{\natexlab{a}}){Krumholz}, {Ellison},
  {Prochaska}, \& {Tumlinson}}]{sfr:kept09}
{Krumholz}, M.~R., {Ellison}, S.~L., {Prochaska}, J.~X., \& {Tumlinson}, J.
  2009{\natexlab{a}}, ArXiv:0906.0983

\bibitem[{{Krumholz} \& {McKee}(2005)}]{sfr:km05}
{Krumholz}, M.~R. \& {McKee}, C.~F. 2005, \apj, 630, 250

\bibitem[{{Krumholz} {et~al.}(2008){Krumholz}, {McKee}, \&
  {Tumlinson}}]{sfr:kmt08}
{Krumholz}, M.~R., {McKee}, C.~F., \& {Tumlinson}, J. 2008, ArXiv:0805.2947,
  805

\bibitem[{{Krumholz} {et~al.}(2009{\natexlab{b}}){Krumholz}, {McKee}, \&
  {Tumlinson}}]{krumholz_etal09}
---. 2009{\natexlab{b}}, \apj, 699, 850

\bibitem[{{Krumholz} \& {Tan}(2007)}]{sfr:kt07}
{Krumholz}, M.~R. \& {Tan}, J.~C. 2007, \apj, 654, 304

\bibitem[{{Mannucci} {et~al.}(2009{\natexlab{a}}){Mannucci}, {Cresci},
  {Maiolino}, {Marconi}, {Pastorini}, {Pozzetti}, {Gnerucci}, {Risaliti},
  {Schneider}, {Lehnert}, \& {Salvati}}]{manucci_etal09}
{Mannucci}, F., {Cresci}, G., {Maiolino}, R., {Marconi}, A., {Pastorini}, G.,
  {Pozzetti}, L., {Gnerucci}, A., {Risaliti}, G., {Schneider}, R., {Lehnert},
  M., \& {Salvati}, M. 2009{\natexlab{a}}, \mnras, 398, 1915

\bibitem[{{Mannucci} {et~al.}(2009{\natexlab{b}}){Mannucci}, {Cresci},
  {Maiolino}, {Marconi}, {Pastorini}, {Pozzetti}, {Gnerucci}, {Risaliti},
  {Schneider}, {Lehnert}, \& {Salvati}}]{hizgal:mcmm09}
---. 2009{\natexlab{b}}, ArXiv:0902.2398

\bibitem[{{Martin} \& {Kennicutt}(2001)}]{sfr:mk01}
{Martin}, C.~L. \& {Kennicutt}, Jr., R.~C. 2001, \apj, 555, 301

\bibitem[{{Mathis} {et~al.}(1983){Mathis}, {Mezger}, \& {Panagia}}]{h2:mmp83}
{Mathis}, J.~S., {Mezger}, P.~G., \& {Panagia}, N. 1983, \aap, 128, 212

\bibitem[{{Mayer} {et~al.}(2008){Mayer}, {Governato}, \&
  {Kaufmann}}]{mayer_etal08}
{Mayer}, L., {Governato}, F., \& {Kaufmann}, T. 2008, Advanced Science Letters,
  1, 7

\bibitem[{{McKee} \& {Krumholz}(2009)}]{mckee_krumholz09}
{McKee}, C.~F. \& {Krumholz}, M.~R. 2009, {\apj} submitted (arXiv/0908.0330)

\bibitem[{{McKee} \& {Ostriker}(2007)}]{sfr:mo07}
{McKee}, C.~F. \& {Ostriker}, E.~C. 2007, \araa, 45, 565

\bibitem[{{Meurer} {et~al.}(1996){Meurer}, {Carignan}, {Beaulieu}, \&
  {Freeman}}]{meurer_etal96}
{Meurer}, G.~R., {Carignan}, C., {Beaulieu}, S.~F., \& {Freeman}, K.~C. 1996,
  \aj, 111, 1551

\bibitem[{{Moster} {et~al.}(2009{\natexlab{a}}){Moster}, {Maccio'},
  {Somerville}, {Johansson}, \& {Naab}}]{moster_etal09b}
{Moster}, B.~P., {Maccio'}, A.~V., {Somerville}, R.~S., {Johansson}, P.~H., \&
  {Naab}, T. 2009{\natexlab{a}}, {\mnras} submitted (arXiv/0906.0764)

\bibitem[{{Moster} {et~al.}(2009{\natexlab{b}}){Moster}, {Somerville},
  {Maulbetsch}, {van den Bosch}, {Maccio'}, {Naab}, \& {Oser}}]{moster_etal09a}
{Moster}, B.~P., {Somerville}, R.~S., {Maulbetsch}, C., {van den Bosch}, F.~C.,
  {Maccio'}, A.~V., {Naab}, T., \& {Oser}, L. 2009{\natexlab{b}}, {\apj}
  submitted (arXiv/0903.4682)

\bibitem[{{Noterdaeme} {et~al.}(2009){Noterdaeme}, {Petitjean}, {Ledoux}, \&
  {Srianand}}]{igm:npls09}
{Noterdaeme}, P., {Petitjean}, P., {Ledoux}, C., \& {Srianand}, R. 2009, \aap,
  505, 1087

\bibitem[{{Ostriker} {et~al.}(2001){Ostriker}, {Stone}, \&
  {Gammie}}]{ostriker_etal01}
{Ostriker}, E.~C., {Stone}, J.~M., \& {Gammie}, C.~F. 2001, \apj, 546, 980

\bibitem[{{Pelupessy} \& {Papadopoulos}(2009)}]{pelupessy_popadopoulos09}
{Pelupessy}, F.~I. \& {Papadopoulos}, P.~P. 2009, {\apj} submitted
  (arxiv/0910.4905)

\bibitem[{{Pelupessy} {et~al.}(2006){Pelupessy}, {Papadopoulos}, \& {van der
  Werf}}]{pelupessy_etal06}
{Pelupessy}, F.~I., {Papadopoulos}, P.~P., \& {van der Werf}, P. 2006, \apj,
  645, 1024

\bibitem[{{Pettini} {et~al.}(2001){Pettini}, {Shapley}, {Steidel}, {Cuby},
  {Dickinson}, {Moorwood}, {Adelberger}, \& {Giavalisco}}]{hizgal:pssc01}
{Pettini}, M., {Shapley}, A.~E., {Steidel}, C.~C., {Cuby}, J.-G., {Dickinson},
  M., {Moorwood}, A.~F.~M., {Adelberger}, K.~L., \& {Giavalisco}, M. 2001,
  \apj, 554, 981

\bibitem[{{Prochaska} {et~al.}(2003){Prochaska}, {Gawiser}, {Wolfe}, {Castro},
  \& {Djorgovski}}]{prochaska_etal03}
{Prochaska}, J.~X., {Gawiser}, E., {Wolfe}, A.~M., {Castro}, S., \&
  {Djorgovski}, S.~G. 2003, \apjl, 595, L9

\bibitem[{{Prochaska} \& {Wolfe}(2009)}]{prochaska_wolfe09}
{Prochaska}, J.~X. \& {Wolfe}, A.~M. 2009, \apj, 696, 1543

\bibitem[{{Rafelski}(2009)}]{rafelski09}
{Rafelski}, M. 2009, private communication

\bibitem[{{Rafelski} {et~al.}(2009){Rafelski}, {Wolfe}, {Cooke}, {Chen},
  {Armandroff}, \& {Wirth}}]{hizgal:rwcc09}
{Rafelski}, M., {Wolfe}, A.~M., {Cooke}, J., {Chen}, H.-W., {Armandroff},
  T.~E., \& {Wirth}, G.~D. 2009, ArXiv:0908.0343

\bibitem[{{Robertson} {et~al.}(2006){Robertson}, {Bullock}, {Cox}, {Di Matteo},
  {Hernquist}, {Springel}, \& {Yoshida}}]{robertson_etal06}
{Robertson}, B., {Bullock}, J.~S., {Cox}, T.~J., {Di Matteo}, T., {Hernquist},
  L., {Springel}, V., \& {Yoshida}, N. 2006, \apj, 645, 986

\bibitem[{{Robertson} {et~al.}(2004){Robertson}, {Yoshida}, {Springel}, \&
  {Hernquist}}]{robertson_etal04}
{Robertson}, B., {Yoshida}, N., {Springel}, V., \& {Hernquist}, L. 2004, \apj,
  606, 32

\bibitem[{{Robertson} \& {Kravtsov}(2008)}]{robertson_kravtsov08}
{Robertson}, B.~E. \& {Kravtsov}, A.~V. 2008, \apj, 680, 1083

\bibitem[{{Roychowdhury} {et~al.}(2009){Roychowdhury}, {Chengalur}, {Begum}, \&
  {Karachentsev}}]{roychowdhury_etal09}
{Roychowdhury}, S., {Chengalur}, J.~N., {Begum}, A., \& {Karachentsev}, I.~D.
  2009, \mnras, 397, 1435

\bibitem[{{Rudd} {et~al.}(2008){Rudd}, {Zentner}, \& {Kravtsov}}]{rudd_etal08}
{Rudd}, D.~H., {Zentner}, A.~R., \& {Kravtsov}, A.~V. 2008, \apj, 672, 19

\bibitem[{{Schaye}(2004)}]{schaye04}
{Schaye}, J. 2004, \apj, 609, 667

\bibitem[{{Schaye} \& {Dalla Vecchia}(2008)}]{schaye_dallavecchia08}
{Schaye}, J. \& {Dalla Vecchia}, C. 2008, \mnras, 383, 1210

\bibitem[{{Schmidt}(1959)}]{schmidt59}
{Schmidt}, M. 1959, \apj, 129, 243

\bibitem[{{Springel} {et~al.}(2006){Springel}, {Frenk}, \&
  {White}}]{cosmo:sfw06}
{Springel}, V., {Frenk}, C.~S., \& {White}, S.~D.~M. 2006, \nat, 440, 1137

\bibitem[{{Springel} \& {Hernquist}(2003)}]{misc:sh03}
{Springel}, V. \& {Hernquist}, L. 2003, \mnras, 339, 312

\bibitem[{{Springel} \& {Hernquist}(2005)}]{springel_hernquist05}
---. 2005, \apjl, 622, L9

\bibitem[{{Stahler} \& {Palla}(2005)}]{stahler_palla05}
{Stahler}, S.~W. \& {Palla}, F. 2005, {The Formation of Stars}

\bibitem[{{Steinmetz} \& {Navarro}(2002)}]{steinmetz_navarro02}
{Steinmetz}, M. \& {Navarro}, J.~F. 2002, New Astronomy, 7, 155

\bibitem[{{Tasker} \& {Bryan}(2008)}]{tasker_bryan08}
{Tasker}, E.~J. \& {Bryan}, G.~L. 2008, \apj, 673, 810

\bibitem[{{Tasker} \& {Tan}(2009)}]{tasker_tan09}
{Tasker}, E.~J. \& {Tan}, J.~C. 2009, \apj, 700, 358

\bibitem[{{Tassis}(2007)}]{tassis07}
{Tassis}, K. 2007, \mnras, 382, 1317

\bibitem[{{Tegmark} {et~al.}(2006){Tegmark}, {Eisenstein}, {Strauss}, \& {et
  al.}}]{tegmark_etal06}
{Tegmark}, M., {Eisenstein}, D.~J., {Strauss}, M.~A., \& {et al.} 2006, \prd,
  74, 123507

\bibitem[{{Wada} {et~al.}(2002){Wada}, {Meurer}, \& {Norman}}]{wada_etal02}
{Wada}, K., {Meurer}, G., \& {Norman}, C.~A. 2002, \apj, 577, 197

\bibitem[{{Wada} \& {Norman}(2001)}]{wada_norman01}
{Wada}, K. \& {Norman}, C.~A. 2001, \apj, 547, 172

\bibitem[{{Wada} \& {Norman}(2007)}]{wada_norman07}
---. 2007, \apj, 660, 276

\bibitem[{{Wetzel} \& {White}(2009)}]{wetzel_white09}
{Wetzel}, A.~R. \& {White}, M. 2009, {\mnras} submitted (arxiv/0907.0702)

\bibitem[{{White} \& {Rees}(1978)}]{white_rees78}
{White}, S.~D.~M. \& {Rees}, M.~J. 1978, \mnras, 183, 341

\bibitem[{{Wild} {et~al.}(2007){Wild}, {Hewett}, \& {Pettini}}]{wild_etal07}
{Wild}, V., {Hewett}, P.~C., \& {Pettini}, M. 2007, \mnras, 374, 292

\bibitem[{{Wolfe} \& {Chen}(2006)}]{sfr:wc06}
{Wolfe}, A.~M. \& {Chen}, H.-W. 2006, \apj, 652, 981

\bibitem[{{Wolfe} {et~al.}(2004){Wolfe}, {Howk}, {Gawiser}, {Prochaska}, \&
  {Lopez}}]{wolfe_etal04}
{Wolfe}, A.~M., {Howk}, J.~C., {Gawiser}, E., {Prochaska}, J.~X., \& {Lopez},
  S. 2004, \apj, 615, 625

\bibitem[{{Wolfe} {et~al.}(2008){Wolfe}, {Prochaska}, {Jorgenson}, \&
  {Rafelski}}]{wolfe_etal08}
{Wolfe}, A.~M., {Prochaska}, J.~X., {Jorgenson}, R.~A., \& {Rafelski}, M. 2008,
  \apj, 681, 881

\bibitem[{{Wolfire} {et~al.}(2003){Wolfire}, {McKee}, {Hollenbach}, \&
  {Tielens}}]{wolfire_etal03}
{Wolfire}, M.~G., {McKee}, C.~F., {Hollenbach}, D., \& {Tielens}, A.~G.~G.~M.
  2003, \apj, 587, 278

\bibitem[{{Wong} \& {Blitz}(2002{\natexlab{a}})}]{misc:wb02}
{Wong}, T. \& {Blitz}, L. 2002{\natexlab{a}}, \apj, 569, 157

\bibitem[{{Wong} \& {Blitz}(2002{\natexlab{b}})}]{wong_blitz02}
---. 2002{\natexlab{b}}, \apj, 569, 157

\end{thebibliography}

\end{document}